\documentclass[preprint,12pt]{elsarticle}

\usepackage{amssymb}
\usepackage{graphicx}
\usepackage{graphics}
\usepackage{rotating}
\usepackage{longtable}
\usepackage{lscape}
\usepackage{epsfig}
\usepackage{rotating}
\usepackage{ifthen}
\usepackage{multicol}
\usepackage{amssymb}
\usepackage{xspace}

\begin{document}

\begin{frontmatter}


\title{Detection of krypton in xenon for dark matter applications}

\author{A.~Dobi$^{a}$}
\author{C.~Davis$^{a}$}
\author{C.~Hall$^{a}$}
\author{T.~Langford$^{a,b}$}
\author{S.~Slutsky$^{a}$}
\author{Y.-R.~Yen$^{a}$}

\address{$^{a}$Department of Physics, University of Maryland, College Park MD, 20742 USA}
\address{$^{b}$Institute for Research in Electronics and Applied Physics, University of Maryland, College Park MD, 20742 USA}


\begin{abstract}
We extend our technique for observing very small 
concentrations of impurities in xenon gas
to the problem of krypton detection. 
We use a conventional mass
spectrometer to identify the krypton content of
the xenon, but we improve the sensitivity 
of the device by more than five orders of magnitude 
with a liquid nitrogen cold trap. 
We find that the absolute krypton concentration in the
xenon can be inferred from the mass spectrometry
measurements, and we identify krypton
signals at concentrations 
as low as  $0.5 \times 10^{-12}$ mol/mol 
(Kr/Xe). This technique simplifies the 
monitoring of krypton backgrounds for WIMP dark 
matter searches in liquid xenon.
\end{abstract}

\end{frontmatter}

\section{Introduction}
\label{intro}

One of the primary challenges faced by liquid 
xenon WIMP dark matter experiments
is the presence of trace amounts of 
radioactive krypton. 
Xenon itself has no long-lived 
radioactive isotopes which might
act as background sources, but krypton 
includes the troublesome anthropogenic isotope 
$^{85}$Kr, a beta emitter with a Q value of 687 keV and a 
half-life of 10.76 years. $^{85}$Kr is created 
in nuclear power plants and released 
into the earth's atmosphere during 
fuel reprocessing, 
and its isotopic fraction at present 
is about $2 \times 10^{-11}$ mol/mol
($^{85}$Kr/$^{nat}$Kr)\cite{kr85_review}.
Xenon, on the other hand, is extracted from the 
atmosphere with a residual krypton concentration 
typically ranging from $10^{-9}$ 
to $10^{-6}$ mol/mol  
($^{nat}$Kr/Xe).
Although this implies that the absolute 
concentration of $^{85}$Kr in xenon is rather small,
the $^{85}$Kr beta decay is nevertheless 
highly problematic for dark matter experiments because
these decays are not suppressed by self-shielding
and because krypton cannot be separated from xenon 
with conventional chemical purifiers. 

The acceptable krypton concentration for a particular experiment is determined by its design sensitivity and by its nuclear recoil discrimination factor. As an example, the LUX dark matter experiment, a dual phase liquid xenon TPC with a recoil discrimination factor of 99.5\%, requires that the residual krypton concentration of the xenon target material be no more than $\sim 3 \times 10^{-12}$ mol/mol ($^{nat}$Kr/Xe)\footnote{All concentrations in this article refer to the natural krypton to xenon ratio, measured in units of mol/mol, unless otherwise indicated.}\cite{lux,lux-simon} in order to be sensitive to a 100 GeV WIMP with a cross section as small as $7 \times 10^{-46}$ cm$^2$. Other liquid xenon detectors, such as XMASS\cite{xmass}, and XENON100\cite{xenon100_first_results, xenon100_backgrounds}, also have demanding krypton goals, and future upgrades of these experiments will require reducing the krypton concentration even further.

To achieve these ultra-low krypton concentrations, commercially procured xenon must undergo additional processing via distillation or gas chromatography. Once this processing is complete, the residual krypton content of the xenon can be determined by low background counting of the $^{85}$Kr beta decays\cite{distillation,xenon100_first_results}, by chromatography\cite{chromotography}, or by atmospheric pressure ionization mass spectroscopy (API-MS)\cite{distillation}\footnote{It has also been suggested that atomic trap trace analysis may be sensitive to krypton at the level of $3 \times 10^{-14}$ mol/mol, but this method has not yet been demonstrated for this application\cite{atta}.}. These methods have achieved a sensitivity of about $\sim 10^{-12}$ mol/mol. Krypton monitoring is useful because it can confirm that the processing has been successful prior to full detector operations, and because it can constrain the background count rate due to $^{85}$Kr in the WIMP search data.

In this article we show that very small krypton 
concentrations can be observed in xenon gas using 
a mass spectrometry  technique which we previously 
developed to detect electronegative impurities in 
xenon\cite{coldtrap}. The method is 
inexpensive, highly sensitive, and it could be quickly adopted and 
applied by many working dark matter experiments. 

\section{Xenon cold trap mass spectrometry}

We use a residual gas analyzer (RGA) 
mass spectrometer to analyze  
our xenon by introducing a 
small quantity of the gas into the 
RGA's vacuum enclosure through a leak valve. 
Since the partial pressure of each
component species is proportional to both 
its absolute concentration 
and to the flow rate through the analysis system, 
by controlling for the flow rate 
the partial pressures can be interpreted in terms 
of the absolute concentrations. The measurement 
is calibrated by preparing 
samples of xenon gas with known impurity concentrations for
the various species of interest by directly mixing known quantities 
of impurities with a known amount of xenon. 
The flow rate can be controlled either by 
measuring the actual flow rate in real time or by
using a standard leak-valve setting whose flow rate 
was previously calibrated.

Since the partial pressures of all species 
are proportional to the flow rate, the RGA signals 
can be vastly increased simply by opening the leak valve further.
However, the RGA cannot be operated above some maximum total 
pressure, typically about $10^{-5}$ Torr, 
and the total pressure is dominated
by the xenon present in the gas sample. 
This limits the maximum flow rate that can be used.
For example, if krypton can be detected by the RGA 
at a partial pressure of $\sim 10^{-12}$ Torr, and
the xenon pressure is $10^{-5}$ Torr, then
the limit of detection is about one part in $10^7$.
Since we are interested in krypton 
concentrations at the level of 
$10^{-12}$ mol/mol, this is 
inadequate for our purposes.

We solve the saturation problem simply 
by removing most of the xenon from 
the gas sample with a liquid nitrogen cold trap placed
between the leak valve and the RGA which allows the flow rate to be vastly 
increased without saturating the RGA.  For the bulk xenon, the pressure is adequate for xenon ice to form ($>$1.8 mTorr at 77 K) \cite{xeice}. So the xenon pressure is held fixed at its vapor pressure. For many common species however, the partial pressure is below the solid-vapor or liquid-vapor equilibrium. This prevents the impurities from becoming trapped.
In our previous paper, we showed that 
impurity species such as oxygen, nitrogen, 
and methane pass through the cold trap in large quantities, 
and that their partial pressures, corrected
for flow rate, remain proportional to their absolute
concentrations. The sensitivity to oxygen, nitrogen, 
and methane was found to be $0.66 \times 10^{-9}$, 
$9.4 \times 10^{-9}$, and $0.49 \times 10^{-9}$ 
(mol/mol), respectively \cite{coldtrap}.

Here we extend the technique to observe krypton 
in xenon. We expect that
krypton could be observed in very small quantities 
by the RGA because there are very 
few background species which could obscure the 
krypton signal. In fact, we find that we are able 
to detect krypton in xenon at a concentration of 
$0.5 \times 10^{-12}$ mol/mol, 
which makes this technique 
better than or comparable to existing methods, 
and sensitive enough to be useful for working
dark matter experiments.

\section{Apparatus and procedures}

\begin{figure}[t!]\centering
\includegraphics[width=100mm]{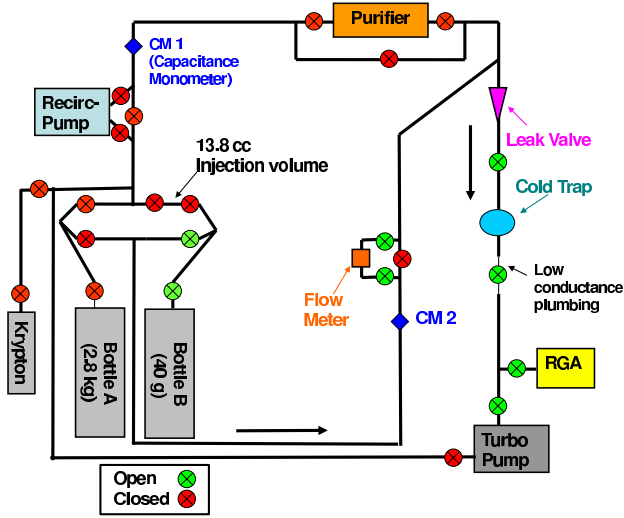}
\caption{Diagram of the xenon handling system. The flow pattern shown is typical of a measurement described in section 5.}
\label{fig:diagram}
\end{figure}

\begin{figure}[t!]\centering
\includegraphics[width=100mm]{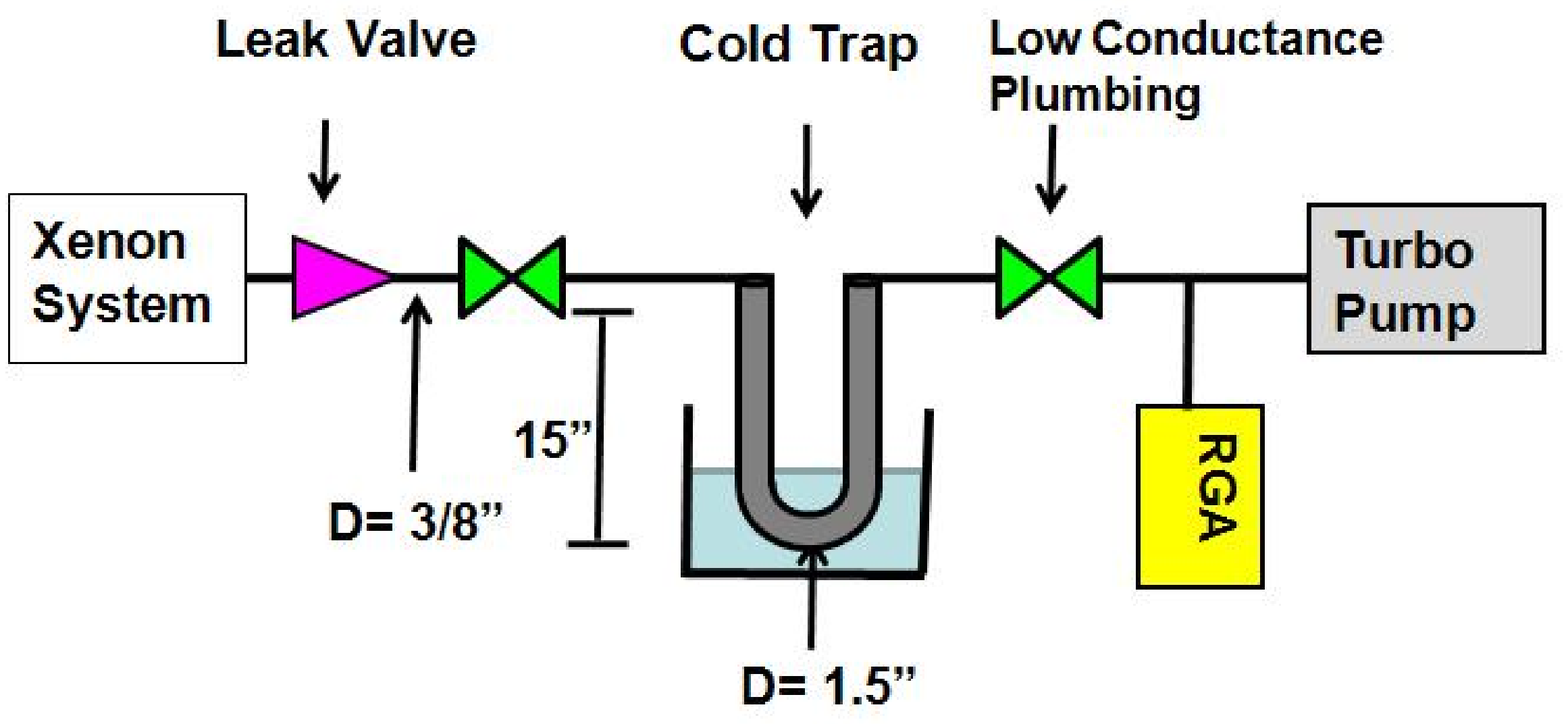}
\caption{Diagram of the cold trap mass spectrometer analysis system.}
\label{fig:coldtrap_diagram}
\end{figure}

A diagram of our analysis apparatus and our xenon handling system
is shown in Figure \ref{fig:diagram} and Figure \ref{fig:coldtrap_diagram}.
The xenon of interest is admitted 
into the analysis system through an 
ultra-high vacuum leak valve (Kurt 
Lesker part number VZLVM940R). It
passes through a liquid nitrogen cold trap 
and a section of low-conductance 
plumbing before reaching an SRS RGA200 
mass spectrometer. 
The low-conductance plumbing is necessary 
to reduce the xenon partial 
pressure from $1.8 \times 10^{-3}$ Torr 
(its vapor pressure at liquid nitrogen
temperature) 
to $< 10^{-5}$ Torr. This 
ensures that the RGA, with the electron multiplier on, remains 
unsaturated. We use a fully open 
hand valve for the low-conductance element.

The cold trap is constructed from 1.5'' 
OD stainless-steel tubing, welded into 
a U shape with a radius of 2.5'' 
and 12.5'' linear inlet and outlet 
legs\footnote{We have also constructed
working cold traps from standard vacuum
plumbing components with 2.75"
CF flanges.}. 
The tubing diameter is chosen to allow
a significant amount of xenon to be analyzed 
before the growth of xenon ice blocks 
the flow of gas through the analysis
system. We find that the liquid nitrogen 
level must be high enough to submerse 
the bottom of the cold trap U, but otherwise 
its level is not critical. 

We use several procedures to calibrate and monitor 
the flow rate through the leak valve. First, 
we calibrate the flow rate directly for a 
variety of leak valve settings using a fixed 
volume of xenon gas ($\sim$ 1 liter) at the 
leak valve input. We measure the pressure 
drop in this volume with a capacitive manometer (Type 627D Baratron)
as the gas flows through the leak valve to 
infer the flow rate for each leak valve setting.
We find that the flow rate is repeatable 
to within 10\% simply by returning the leak 
valve to the same indicator marking on its dial. 
Second, we also measure the flow 
rate directly using a MKS model 179A 
mass flow meter which has been 
calibrated for use with xenon gas.
Third, when the xenon under analysis 
contains a small, constant concentration of 
a tracer gas which is unaffected by the 
cold trap, such as argon or helium, then the 
partial pressure of the tracer can 
be used to accurately monitor the leak rate 
in real time with the RGA itself. This 
eliminates the systematic error due to 
the leak valve dial setting and RGA gain
drift. We use this method in 
Section \ref{sec:constant-pressure}.

To calibrate the partial pressure measurements 
of the RGA in terms of the true krypton 
concentration, we insert known quantities of
krypton into the xenon using a krypton gas 
cylinder (99.999\% krypton purity). The
injection volume is $13.8\pm 0.1$ cc of plumbing 
monitored by a pressure gauge, accurate within 0.1 Torr, and isolated 
by two valves. We inject krypton with pressures above 20 Torr, and further reduce the pressure by volume sharing. The
krypton is combined with the xenon 
by flowing the xenon through 
the injection volume and collecting the 
gases in a recovery bottle where they mix.

To perform a measurement, first we submerge 
the cold trap in liquid nitrogen while it 
is pumped to ultra-high vacuum by the 
turbo-molecular pump, typically reaching a vacuum of $4\times 10^{-8}$ Torr. We then 
open the leak valve in two steps. In the 
first step, we use a very small leak rate, 
less than $10^{-4}$ standard liters per minute
(SLPM), which allows xenon ice 
to form in the cold trap, establishes 
the fixed xenon partial pressure, and flushes 
some trace background gases out of
the analysis system plumbing. We wait for 
several minutes for the partial pressures 
of all species to stabilize, and then we 
open the leak valve to the desired flow 
rate for purity analysis. In general, the 
best sensitivity is obtained by using the 
maximum possible flow rate. 
In some cases the flow rate is 
limited by the partial pressure of 
non-xenon impurity species such as 
oxygen, nitrogen, or argon. Since these 
species are not removed by the cold trap, 
their presence in the xenon gas will eventually
cause the RGA to saturate as the flow 
rate is increased. For the very best 
sensitivity, the xenon should be free 
from extraneous impurity species.

\section{Response of the analysis system to krypton}
\label{sec:constant-pressure}

In our first series of experiments, we confirm 
that the krypton partial pressure 
observed by the RGA is indeed linear in the 
true concentration by injecting known amounts 
of krypton into our 2.8 kg xenon supply. For these
measurements, the xenon supply bottle continuously 
feeds xenon into the system through a regulator, 
maintaining a constant pressure at the leak valve 
input. This insures that the leak rate into the 
analysis system is nearly constant throughout the 
measurement, which simplifies the data analysis.

To precisely monitor the leak valve flow rate in 
real time, we use argon as a tracer gas. Our 
2.8 kg xenon supply contains an argon 
concentration of about $10^{-6}$ mol/mol, 
and since the argon level is constant 
from one injection experiment to the next (because 
the injected gas is 99.999\% krypton, with only 
trace quantities of argon), the argon partial 
pressure serves as a convenient proxy for the gas 
flow rate through the analysis system. Under these
conditions, the leak valve can be opened to an 
arbitrary setting, and the krypton-to-argon 
partial pressure ratio should be proportional 
to the true krypton concentration.

\begin{figure}[t!]\centering
\includegraphics[width=100mm]{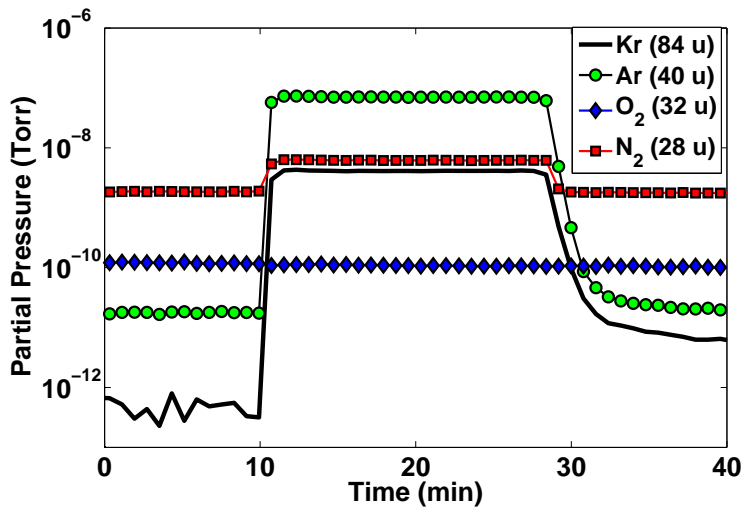}
\caption{The results for a typical purity 
measurement with constant 
flow rates. 
At  t = 10 minutes the leak valve is opened and a 
measurement is made with a flow rate of 0.10 SLPM. 
This sample of xenon contained 
$40 \times 10^{-9}$ mol/mol nitrogen, 
$4 \times 10^{-6}$ mol/mol argon, and $66.4 \times 10^{-9}$
mol/mol krypton. At t = 28 minutes 
the leak valve is closed.}
\label{fig:ConstantFlow}
\end{figure}

A typical measurement is shown in Figure 
\ref{fig:ConstantFlow}. At t = 0, xenon ice 
has already been established in the cold trap and backgrounds have stabilized, 
and the measurement begins at t = 10 minutes, with a flowrate of 0.10 SLPM. 
Krypton, argon, and nitrogen are clearly 
present in the sampled gas, while the
oxygen concentration is less than 
$0.7 \times 10^{-9}$ mol/mol. 
At t = 28 minutes the leak valve is closed.

\begin{table}[t!]
\begin{centering}
\begin{tabular}{|c|c|c|c|}
\hline
$\Delta \rho$(Kr) & $P_{Kr}$ (84 u) & 
                     $P_{Ar}$ (40 u) & $P_{Kr}/P_{Ar} $  \\ 
 ($10^{-9}$ mol/mol) & ($10^{-9}$ Torr) 
  & ($10^{-9}$ Torr) &  (Torr/Torr) \\ 
\hline
0     & 4.11 & 70.3 & 0.0584 \\ \hline
7.37  & 5.90 & 91.2 & 0.0647 \\ 
      & 6.02 & 91.9 & 0.0655 \\ \hline
18.4 & 7.00 & 93.7 & 0.0747 \\ 
      & 6.60 & 87.8 & 0.0752 \\ \hline
33.1 & 7.88 & 89.6 & 0.0879 \\ 
      & 7.51 & 86.0 & 0.0873 \\ 
      & 8.05 & 92.0 & 0.0876 \\ 
      & 7.76 & 89.1 & 0.0871 \\ \hline
\end{tabular}
\caption{Krypton and argon partial pressures as a function 
of the injected krypton concentration ($\Delta \rho$(Kr)). 
Each gas sample was measured at least twice to 
confirm the repeatability of the Kr-to-Ar ratio. The uncertainty is the amount of krypton injected depends on the error in the injection volume and the error in the pressure gauge, combined they are less than 1\%. The partial pressure recorded by the RGA is averaged during the measurement yielding a statistical uncertainty of 1\% in partial pressure.}
\label{tab:natural_xe_tab}
\end{centering}
\end{table}

\begin{figure}[t!]\centering
\includegraphics[width=100mm]{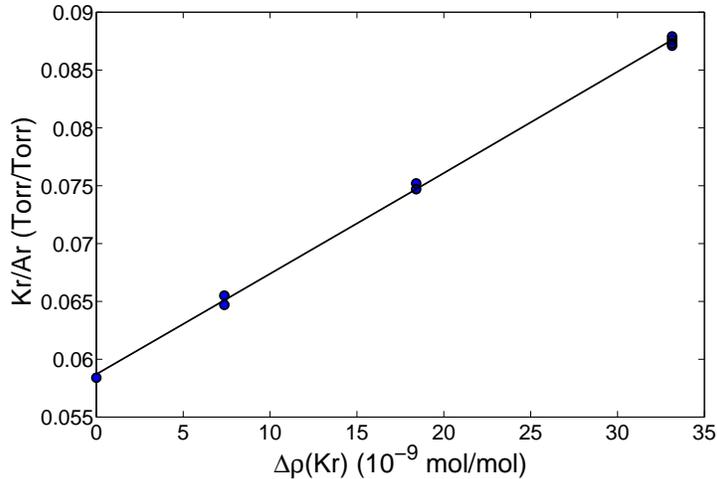}
\caption{Krypton-to-argon partial pressure
ratio versus the injected krypton concentration
 ($\Delta \rho$(Kr)). Repeated measurements are
shown as separate data points.
The non-zero y-intercept value is due to the 
krypton present in our vendor supplied xenon
before our injections. Each gas sample was measured at least twice to 
gauge the systematic uncertainty in the RGA's partial pressure
measurements for a fixed concentration of krypton. 
We infer an uncertainty of 1\% in the krypton-to-argon  
partial pressure ratio for a fixed concentration of krypton. Error bars are not plotted as they are too small to be seen on the graph}
\label{fig:natural_xe}
\end{figure}

In Figure \ref{fig:natural_xe} 
and Table \ref{tab:natural_xe_tab} we show the 
krypton-to-argon partial pressure ratio as 
a function of the amount of krypton which 
we inject into our xenon supply. The partial pressure recorded by the RGA is averaged during the measurement yielding a statistical uncertainty of less than 1\%. The uncertainty in the krypton to xenon ratio after an injection is 1\%, from the uncertainty in the xenon mass (2800$\pm 20$g), the injection volume, and the error on the pressure gauge.  
As shown in Table \ref{tab:natural_xe_tab} each gas sample was measured at least twice to 
study the repeatability of the krypton-to-argon ratio for a fixed purity concentration. We find the ratio repeatable to about 1\%. The absolute concentration of the argon is known only to within 50\% ($1\pm 0.5$ ppm), however, the absolute concentration is not relevant, since we only use the argon as a flow rate standard. 

The krypton-to-argon ratio 
is found to be linear in the injected concentration, 
which confirms that the cold trap allows the 
krypton to pass through as desired. Additional data confirms that it is also linear in the flow rate as observed for other species \cite{coldtrap}. In total 
we injected $33.1 \times 10^{-9}$ mol/mol  
of krypton, which 
resulted in a total increase in the 
krypton-to-argon ratio of a factor of 
1.50  relative to the 
vendor-supplied xenon. From this we infer 
that the krypton concentration 
was $(66.4\pm 4) \times 10^{-9}$ mol/mol
before our injections, and 
$(99.5\pm 4) \times 10^{-9}$
mol/mol after injections.

\section{Detection of krypton at the $10^{-12}$ mol/mol level}

The measurements described in Section \ref{sec:constant-pressure} allow us to quantify the response of the analysis system to krypton. For example, the data in Table \ref{tab:natural_xe_tab} shows a krypton partial pressure of $7.9\times 10^{-9}$ Torr, for a flow rate of 0.1 SLPM, and a krypton concentration of $99.5\times 10^{-9}$ mol/mol. Therefore, the analysis system response to krypton is 
$0.79$ Torr/(SLPM $\cdot$ mol/mol).  
Since the fluctuations in the RGA reading 
at 84 AMU (Atomic Mass Unit) are  $\sim 3 \times 10^{-13}$ Torr, 
we expect that a concentration 
of $1 \times 10^{-12}$ mol/mol
krypton could be detectable at a 
flow rate of $\sim 0.4$ SLPM. However, as shown 
in Figure \ref{fig:ConstantFlow}, our xenon 
supply contains significant argon and nitrogen.
Since these trace gases are not removed by 
the cold trap, they will cause the RGA to 
saturate at flow rates above 0.1 SLPM, 
with the argon being the leading problem. This would
limit our krypton sensitivity 
to about $4 \times 10^{-12}$ mol/mol.  
To detect krypton at lower concentrations 
it is necessary to remove these trace impurities. 

Nitrogen can be removed from xenon using standard
getters \cite{purifier}, but argon cannot.
We first tried to remove the argon
by freezing the xenon in its supply bottle 
with liquid nitrogen and pumping on the vapor with
the turbo-molecular pump. This strategy proved
to be inefficient, probably 
because the argon is trapped in the xenon ice requiring long diffusion times to escape. 
However, we successfully purified a small quantity
of xenon using the cold trap itself. We allowed approximately seven standard
liters of xenon to slowly leak 
into the cold trap while continuously pumping with a turbo-pump.
We repeated this process several times and during each cycle we measured the remaining krypton concentration with the RGA. We found that the krypton and argon levels were significantly reduced after each pass. We guess that the reduction comes about because the turbo pump is able to remove the krypton and argon before they become permanently trapped in the slowly forming xenon ice.

\begin{figure}[t!]\centering
\includegraphics[width=100mm]{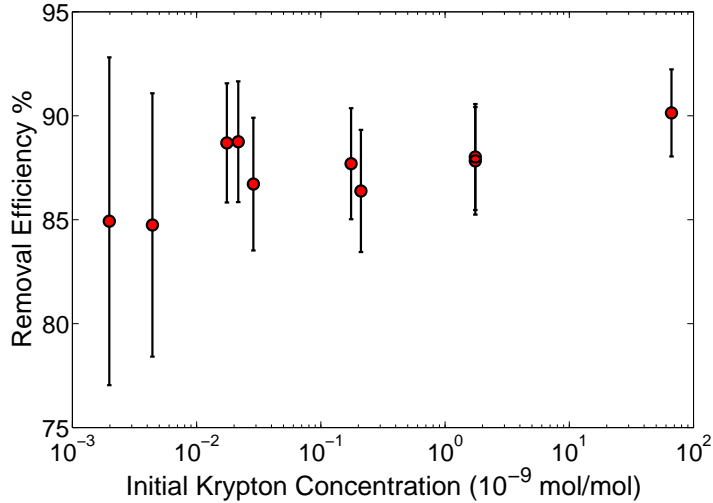}
\caption{One pass krypton removal efficiency of the coldtrap vs. initial concentration, for several purification cycles.}
\label{fig:KrRemoval}
\end{figure}

\begin{figure}[t!]\centering
\includegraphics[width=100mm]{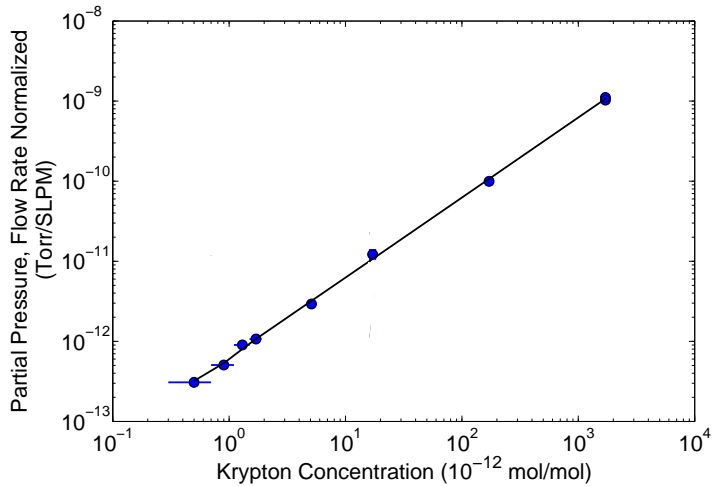}
\caption{Krypton detection figure-of-merit
as a function of true krypton concentration. 
The krypton  partial pressure 
data (at 84 AMU) is averaged in a 60 second 
window around its peak value and
then normalized to the average 
flow rate. The maximum deviation from a 
linear fit over three orders of magnitude 
is  13.7\%. The solid line indicates the linear fit to the data.}
\label{fig:oneppt}
\end{figure}

The average argon and krypton one-pass purification efficiencies were determined to be 99.4\% and 87\% respectively. We find that the krypton purification efficiency remains roughly constant vs. initial concentration over five orders of magnitude, shown in Figure \ref{fig:KrRemoval}. The removal efficiencies are derived assuming that the RGA's response to krypton is linear in concentration, which was later confirmed (see Figure \ref{fig:oneppt}). The uncertainty in the purification efficiency is derived from the maximum deviation from the linear response reported in Table \ref{tab:oneppt}. We do not include the argon removal efficiency in Figure \ref{fig:KrRemoval} because we only have one measurement for argon. After the second pass, argon was not detectable by the RGA due to interference from doubly ionized krypton. This procedure ultimately produced a 40 gram sample of xenon with much 
less than $1 \times 10^{-12}$ mol/mol argon 
and krypton.

Starting with this 40 gram sample of 
de-argonated and de-kryptonated xenon gas, 
we created xenon with known 
krypton concentration by mixing it with 
small quantities of the 
2.8 kg xenon supply, 
which contained $99.5 \times 
10^{-9}$ mol/mol of krypton after the
experiments described in
Section \ref{sec:constant-pressure}. 
For example, to achieve 
$0.5 \times 10^{-12}$ mol/mol of 
krypton, we added 0.19 milligrams of our 
krypton-rich xenon supply to the 
40 grams of de-kryptonated xenon. The sample was created by filling a volume of  $13.8\pm 0.1$ cc with $215.9\pm 0.1$ Torr of xenon containing $(99.5\pm 4) \times 10^{-9}$ mol/mol of krypton and then volume sharing the gas with a $1.5222\pm 0.0005$ L volume to further reduce the pressure. With this simple method xenon samples could be prepared with krypton concentrations known to within 5\% of the injection amount.
The xenon was then analyzed
by the cold trap mass spectrometry 
technique. Between each run we removed 
the extra krypton, and a new sample was mixed 
starting again with de-kryptonated xenon.

\begin{figure}[t!]\centering
\includegraphics[width=60mm]{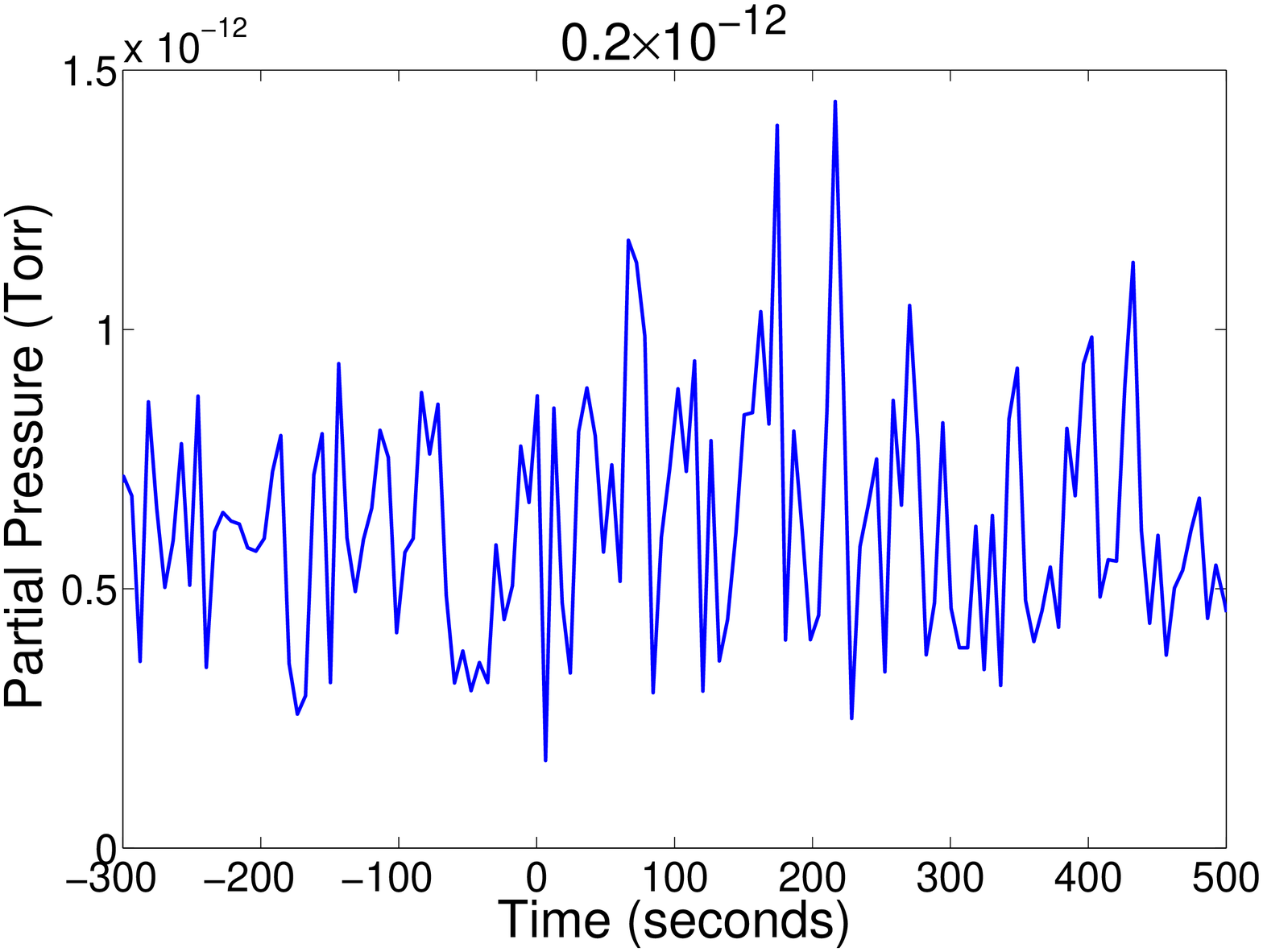}
\includegraphics[width=60mm]{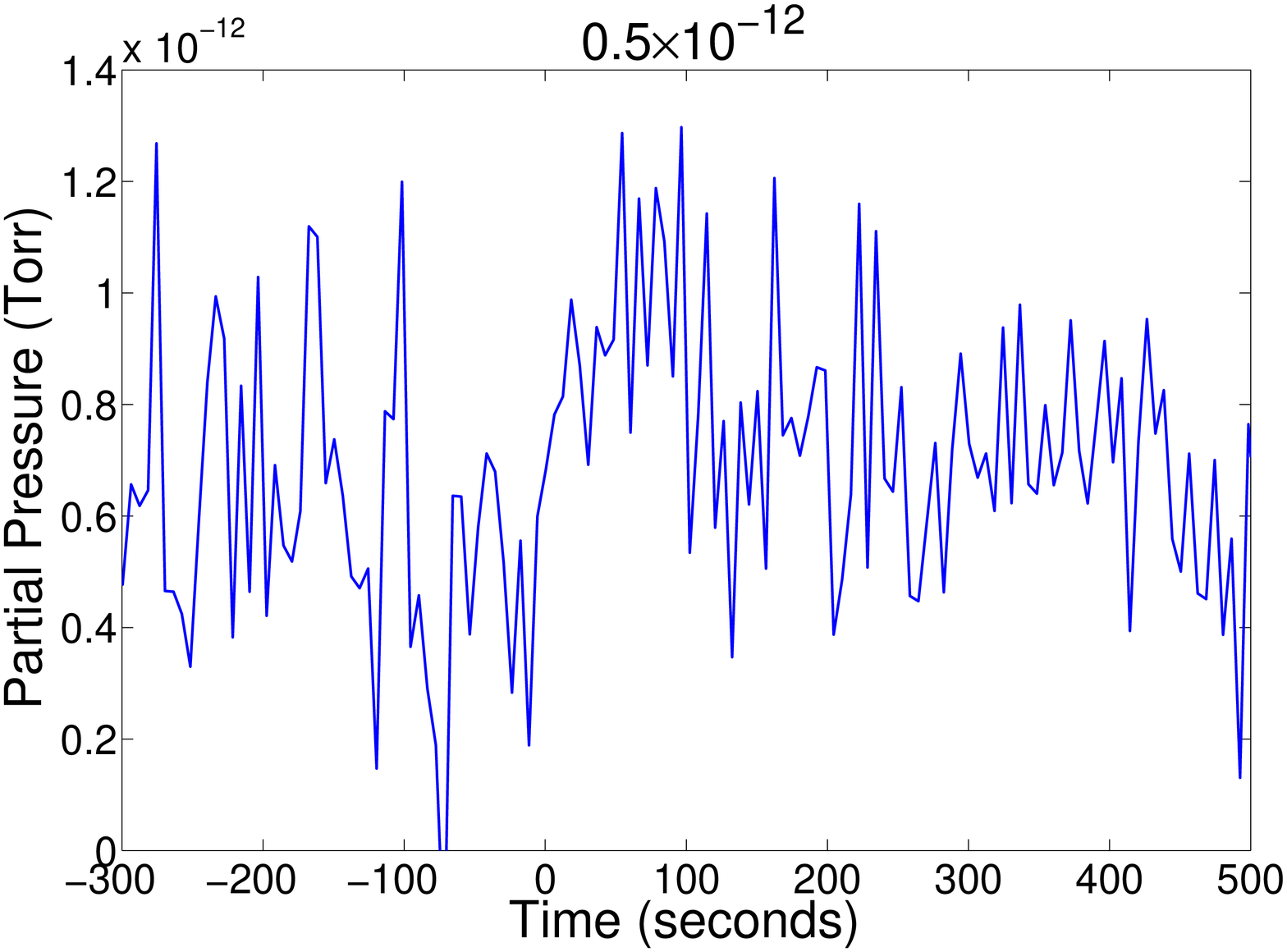}
\includegraphics[width=60mm]{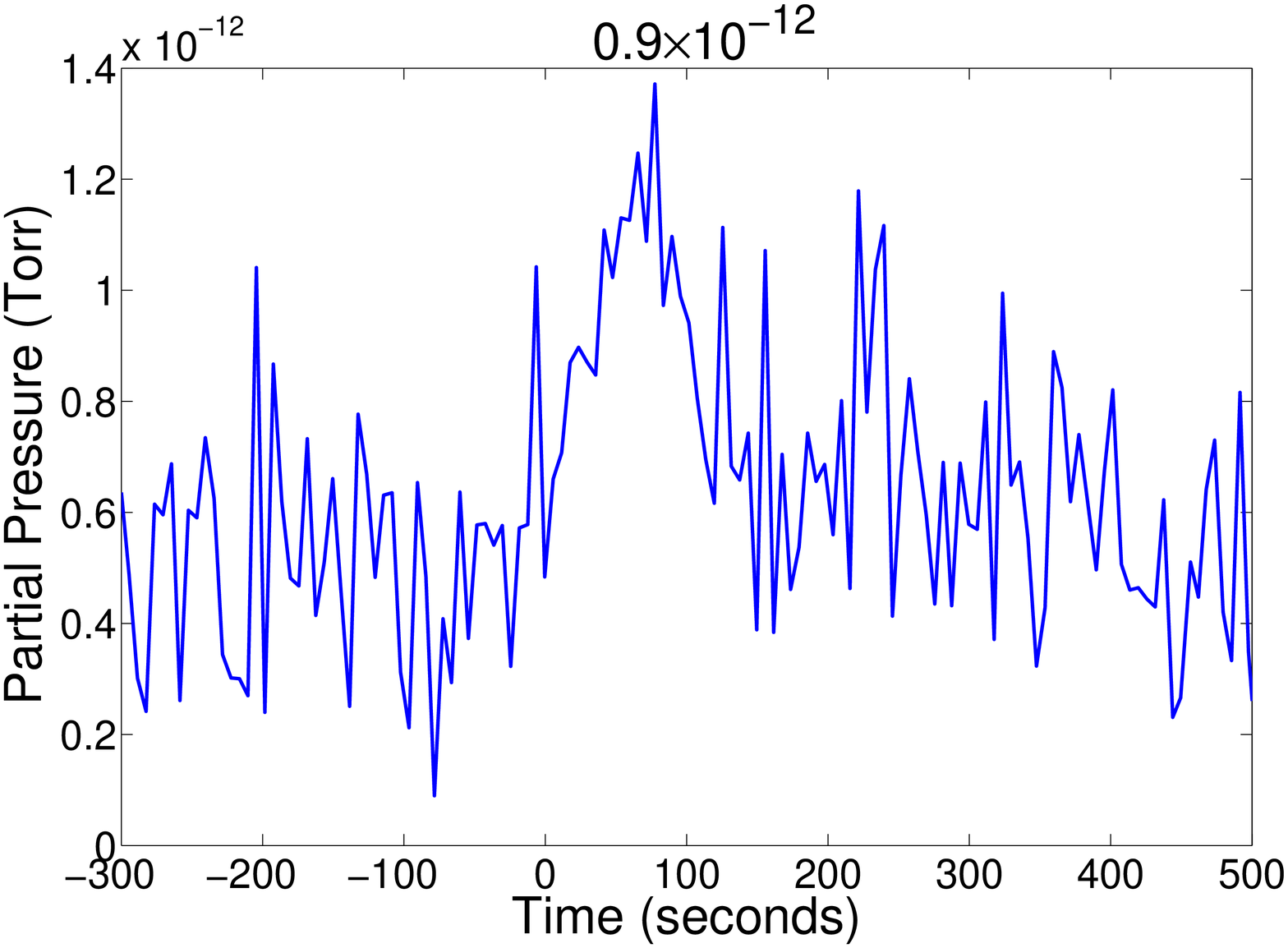}
\includegraphics[width=60mm]{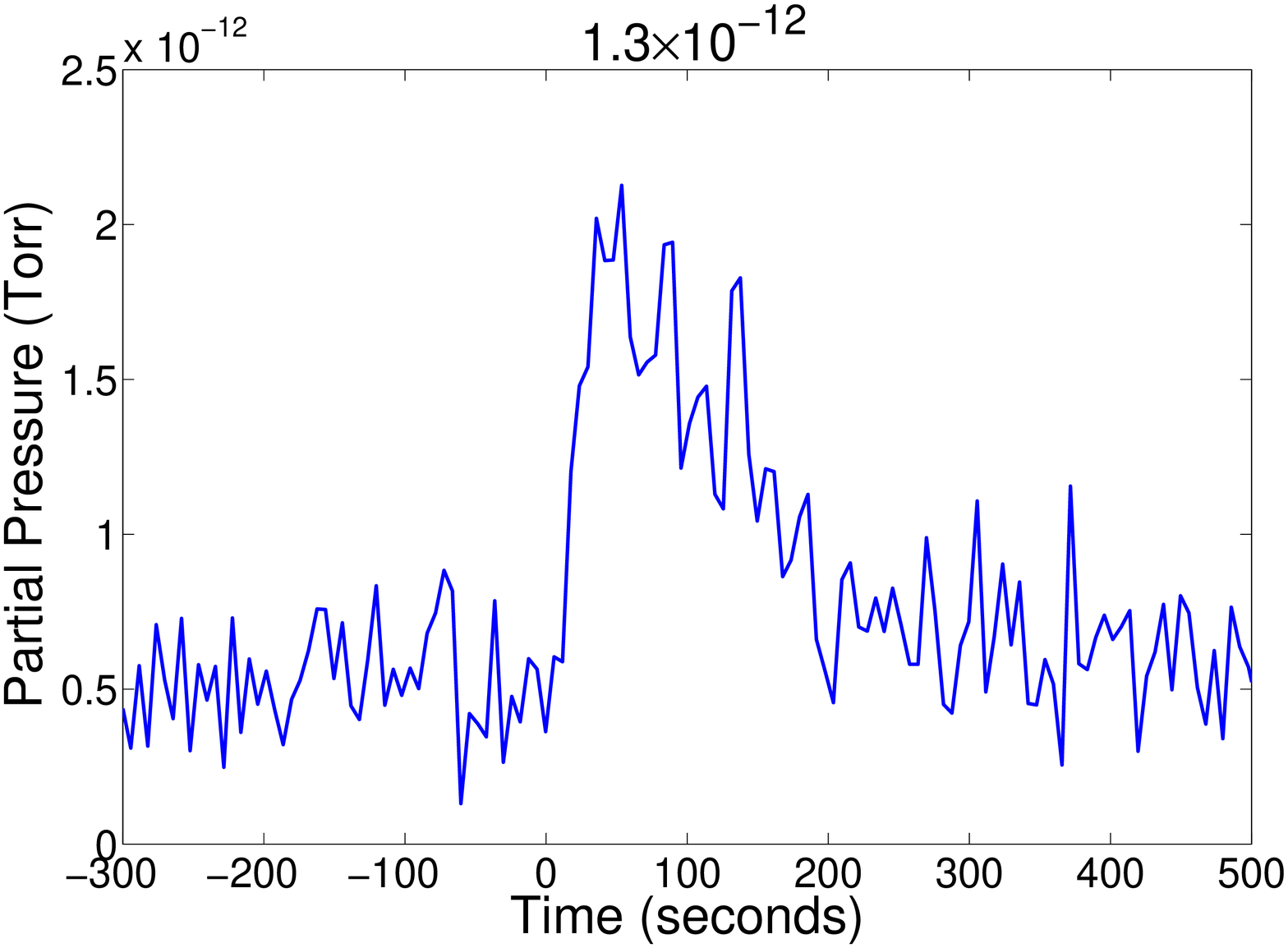}
\caption{RGA response to the smallest 
concentrations of krypton. The krypton 
signals decay in time because the 
flow rate is decreasing due to the small
amount of xenon available for these 
measurements. Upper left: 
$0.2 \times 10^{-12}$ mol/mol. Upper right:
$0.5 \times 10^{-12}$ mol/mol. Lower left:
$0.9 \times 10^{-12}$ mol/mol. Lower right:
$1.3 \times 10^{-12}$ mol/mol. }
\label{fig:peakplots}
\end{figure}

RGA partial pressure plots are shown in 
Figure \ref{fig:peakplots} for xenon with 0.2, 0.5, 
0.9, and 1.3 $\times 10^{-12}$ mol/mol of krypton. 
For the data
shown in these plots we open the 
leak valve to its maximum setting at 
t = 0 seconds. Since the total amount of xenon 
available for these measurements is
modest, the flow rate immediately peaks at 1.5 SLPM 
and then decreases during the
measurement due to the decreasing pressure
at the leak valve input. The resulting
partial pressure plots follow this pattern, 
as shown in Figure \ref{fig:peakplots}.

\begin{table}[t!]
\begin{centering}
\begin{tabular}{|c|c|c|c|c|}
\hline
$\rho$(Kr) & Avg. $P_{Kr}$ & Avg. Flow  & Kr Fig. of Merit &  Dev. from Fit \\ 
($10^{-12}$ mol/mol)  & ($10^{-12}$ Torr) & (SLPM) & ($10^{-12}$ Torr/SLPM) & (\%) \\ \hline
$0.5\pm 0.2$   & 0.406 & 1.32 & $0.308\pm  0.0154$ & 3.8 \\ \hline
$0.9\pm 0.2$   & 0.646 & 1.27 & $0.507\pm 0.0254$  & 5.1 \\ \hline
$1.3\pm 0.2$   & 1.15  & 1.28 & $0.901\pm  0.0450$ & -12.4 \\ \hline
$1.7\pm 0.2$   & 1.37  & 1.28 & $1.07 \pm 0.0528$  & 1.2 \\ \hline
$5.1\pm 0.3$   & 3.81  & 1.30 & $2.93 \pm 0.147$   & 8.6 \\ \hline
$17.1\pm 0.9$  & 15.7  & 1.29 & $12.2 \pm 0.608$   & -13.7 \\ \hline
$171.1\pm 8.8$ & 127   & 1.27 & $99.5 \pm 4.97$    & 6.9 \\ \hline
$^*1711\pm 88$ & 1450  & 1.31 & $1110 \pm  55.3$   & -3.5 \\ 
$1711\pm 88$   & 1330  & 1.29 & $1030 \pm 51.3$    & 4.1 \\ \hline
\end{tabular}
\caption{Results of krypton detection experiments with 
40 grams of highly purified xenon. The
various krypton concentrations 
were created by mixing with xenon containing 
$99.5 \times 10^{-9}$ mol/mol krypton, except the
sample labeled ($^*$), which was created by injecting 
99.999\% krypton from a krypton gas cylinder. 
For prepared samples of $1.7 \times 10^{-12}$ mol/mol 
or less the uncertainty in the concentration is 
dominated by the minimum sensitivity 
to krypton in the highly purified xenon, which we take 
to be $0.2 \times 10^{-12}$ mol/mol. 
For concentrations above $1.7 \times 10^{-12}$ 
the uncertainty in the concentration is 5\%, 
dominated by the uncertainty of the concentration of the  
krypton-rich xenon supply.  
The krypton figure of merit is the average partial 
pressure divided by the average flow rate.
The last column shows the deviation from the 
linear fit shown in Figure \ref{fig:oneppt}.}
\label{tab:oneppt}
\end{centering}
\end{table}

Clear krypton signals are seen for concentrations
of $0.9  \times 10^{-12}$ mol/mol and larger, while 
the $0.5 \times 10^{-12}$ mol/mol sample
gives a marginal signal. No significant deviation 
from background is seen for the 
$0.2 \times 10^{-12}$ mol/mol sample, which  
agrees with our expected sensitivity of
$0.25 \times 10^{-12}$ mol/mol (inferred by
assuming partial pressure fluctuations of  
$3 \times 10^{-13}$ Torr and a 
1.5 SLPM flow rate.)

To quantify the krypton concentration, we 
calculate the average partial pressure in a 
60 second window around its maximal value
after subtracting the background level, 
and we divide by the average flow rate measured
by the MKS mass flow meter. Since the 
same leak valve setting (fully open) was
used in each dataset, the average flow rate
varies by less than 2\% in all datasets.  
As shown in Figure \ref{fig:oneppt} and 
Table \ref{tab:oneppt}, the 
ratio of average pressure to average flow rate 
is proportional 
to the true krypton concentration over four 
orders of magnitude. The linear dependence was expected as the peak partial pressure, for a fixed flow rate, should be proportional to the number of particles passing by the RGA per unit time. The largest deviation 
from the fitted line is 13.7\%, and the 
dataset at $0.5 \times 10^{-12}$ mol/mol 
deviates from the fitted line by only 4\%. 
This indicates that the small krypton
signal at $0.5 \times 10^{-12}$ mol/mol 
is likely to be genuine, and demonstrates
sensitivity to krypton at concentrations 
less than $1 \times 10^{-12}$. Previous methods have achieved sensitivities of about $\sim 10^{-12}$ mol/mol \cite{distillation,chromotography}.

The linear dependence of the krypton figure-of-merit to concentration down to $0.5 \times 10^{-12}$ mol/mol also demonstrates that the de-kryptonated xenon truly had an initial concentration less than $0.5 \times 10^{-12}$ mol/mol. Had there been residual krypton in the xenon before the krypton injections, the data in Figure \ref{fig:oneppt} would level off and not continue its linear decline at lower concentrations.
 
To confirm that the absolute krypton 
concentration of our experiments 
is not in error, we performed one final
krypton injection from the krypton cylinder
into our 40 grams of de-kryptonated xenon 
at a concentration of 
$1.7 \times 10^{-9}$ mol/mol. This dataset
gives a krypton figure of merit which agrees
with the equivalent dataset produced by mixing
to within 8\%. 

A second cross-check on our krypton concentration scale is provided by
the EXO-200 double beta decay experiment. As reported in Ref.
\cite{steveherrin}, EXO-200 has observed $^{85}$Kr in its natural
(unenriched) xenon gas supply at a decay rate consistent with that
inferred from our mass spectrometry technique, assuming that the
$^{85}$Kr isotopic abundance is $\sim 10^{-11}$, as expected. This
confirms our absolute scale to within a factor of two.

\section{Conclusion}

We have extended the xenon cold trap mass spectrometry 
technique to detect trace quantities of krypton in 
xenon gas. We find that krypton passes through the 
cold trap largely undisturbed, and that the 
resulting partial pressure is proportional to 
the true concentration after accounting for the flow rate.
Using this method we have detected krypton concentrations
as low as $(0.5\pm 0.2) \times 
10^{-12}$ mol/mol $^{nat}$Kr/Xe. 

Compared to the previously reported methods for krypton detection in
xenon, our technique is rather simple and inexpensive, yet it achieves
better sensitivity.  It does not require any specialized equipment beyond an RGA, which most labs use routinely anyway.

We believe our sensitivity
could be significantly improved by using faster flow rates,
which could be achieved by using a larger sample of 
highly purified xenon. In principle we see no reason 
why an additional factor of ten improvement could 
not be achieved by using ten times the amount of xenon, 
which would make the technique useful for future
WIMP dark matter experiments. In any case, the sensitivity 
demonstrated here will already be useful for krypton 
monitoring programs at existing detectors.

\section{Acknowledgments}
\label{sec:Acknowledgments}
This work was supported by the National Science Foundation 
under award number PHY0810495.

\bibliographystyle{elsarticle-num}
\bibliography{kr}

\end{document}